\documentclass{emulateapj}

\newcommand{\xmm}{{\sl XMM-Newton }}

\newcommand{\ros}{{\sl ROSAT }}

\newcommand{\eino}{{\sl Einstein Observatory}}

\newcommand{\chandra}{{\sl Chandra }}
\newcommand{\ergsec}{\hbox{erg s$^{-1}$ }}

\newcommand{\asca}{{\sl ASCA }}

\newcommand{\hdstar}{HD~119682}
\newcommand{\wga}{1WGA~J1346.5$-$6255}
\newcommand{\gton}{G309.2$-$0.6}
\newcommand{\magell}{{\sl Magellan }}

\begin{document}

\title{
Extraordinarily Hot X-ray Emission \\
from the O9 Emission Line Star \hdstar\ 
}

\author{Cara E. Rakowski\altaffilmark{1}}
\email{crakowski@cfa.harvard.edu}

\author{N. S. Schulz\altaffilmark{2}}

\author{S. J. Wolk\altaffilmark{1}}


\author{Paola Testa\altaffilmark{2}}

\altaffiltext{1}{Harvard-Smithsonian Center for Astrophysics, 60
  Garden Street Cambridge, MA 02138}
\altaffiltext{2}{Massachusetts Institute of Technology, Center for
  Space Research, 70 Vassar Street, Cambridge, MA 02139}  

\begin{abstract}
We present new optical and X-ray observations to show that the
X-ray source \wga\ previously associated with the SNR
G309.2-0.6 can be unequivocally identified with the emission line
star \hdstar\ located in the foreground open cluster NGC 5281. 
Images from \chandra in the X-ray band as well as from \magell in the 
narrow optical H$\alpha$ band show a coincidence of the source positions
within 0.5\arcsec. The X-ray source appears extremely hot for an OB-star
identified as of O9.7e type. \xmm spectra show plasma temperatures of 
1 keV and $>$ 8 keV with an X-ray luminosity of 
6.2$\pm0.1\times10^{32}$ \ergsec. 
The optical and X-ray properties are very reminiscent 
of the prototype emission line star $\gamma$ Cas.
We discuss the ramifications of this similarity with respect to very
early type emission line stars as a new class
of hard X-ray sources. 
 
\end{abstract}


\section{Introduction}

X-ray emission from early type stars has many origins and much
progress has been made, particularly in recent years, to understand
some of the underlying mechanisms. 
Since the discovery with the
\eino~\citep{harnden1979, seward1979} it has been established that
O-stars and most early-type B-stars emit X-rays of some
form~\citep{pallavicini1981,chlebowski1989,
  berghoefer1994,cassinelli1994}. Observations with \ros established a
fairly tight relationship between the X-ray and bolometric luminosity
with $L_X/L_{bol} \sim 10^{-7}$~\citep{berghoefer1996}. OB-stars that
follow this relation are thought to generate X-rays from a
distribution of shocks in a line driven
wind\citep{lucy1980,lucy1982,owocki1988}. We have a
relatively fair understanding how the shocks lead to X-ray emission
and for involved wind velocities one can derive X-ray emissivities
from plasma temperatures of up to 25 MK. However, once it comes to
processes that lead to these shocks, models differ quite
substantially~\citep{owocki1988, feldmeier1995, howk2000, feldmeier2002}. 

Soon after the launch of \chandra and \xmm it became clear that the
origins of X-rays in hot stars is much more complex and puzzling. Some
early observations confirmed the picture where X-rays arise from shock
fragments embedded within about 10 stellar radii of the stellar
wind~\citep{kahn2001,waldron2001,cassinelli2001,miller2002}, but others
showed unusually high plasma temperatures and X-ray line properties
which seemed to be more related to forms of magnetic activity
~\citep{schulz2000,waldron2001,cohen2003,schulz2003,gagne2005,schulz2006}.
Hard X-rays from hot stars have now been observed from various
circumstellar environments. Enhanced X-ray emission from
magnetically confined winds has been observed in young OB-stars in
and near the Orion Trapezium~\citep{schulz2003,gagne2005,schulz2006} and also
seems the likely process for $\tau$ Sco~\citep{cohen2003, donati2006}. 
Candidates for X-ray
enhancement through magnetic fields or colliding winds 
may be located in other young embedded clusters~\citep{wolk2002,
rho2004, townsleyetal2003}. Hard X-rays from colliding
wind binaries have been observed in several exotic O-star binaries
~\citep{sana2004,corcoran2005,pollock2005}.
Binary interactions can also lead to
powerful hard X-ray outbursts when two magnetically active stars
interact as has been recently proposed for the massive spectroscopic
binary in $\theta^2$ Ori A~\citep{schulz2006}. 

A third category has recently been identified as a possible
source of intrinsic hard X-rays: hot emission line stars. The
prototype of such a source is $\gamma$ Cas, which before \chandra was
known as a hard X-ray source but was considered likely to be a binary
system of a B0.5e star orbited by a degenerate companion, probably
a white dwarf~\citep{kubo1998, owens1999}. \citet{smith2004} recently
reported on high resolution spectra obtained with \chandra which
indicated quite a number of peculiar X-ray properties.
The spectra featured various
emission line spectral components from hot collisional plasmas which
range from 100 eV to 12 keV as well as cold fluorescence. In their
analysis~\citet{smith2004} conclude that the X-ray emission is likely
to originate close to the Be star and its disk.

In this paper we report on X-ray observations of another early-type
star with bright, hot X-ray emission, the O9e emission line star
\hdstar, located in the open cluster NGC~5281. 
The open cluster itself is fairly well-studied.
Photometry and proper motion of NGC~5281  were
investigated by \citet{sanner2001}, who found an age for the
cluster of 45$\pm$10~Myr and a distance of 1580$\pm$150~pc from fits to the
color-magnitude diagram after statistical field star subtraction.
They find a membership probability $P = 0.78$ for \hdstar\ based
on the proper motion. 
\citet{levenhagen2004} spectroscopically determined the
fundamental 
parameters for \hdstar , an effective temperature of $31910 \pm 550$~K, 
an age of $4 \pm 1$~Myr, and 
a luminosity  $log(L/L_{\sun}) = 4.64 \pm 0.1$ implying a distance
modulus consistent with NGC 5281.
Given updated determinations of O star effective
temperatures which account for line-blanketing \citep{martins2002},
this effective temperature categorizes \hdstar\ as an O9.7 star, but
it is certainly on the borderline between O and B, with
\citet{levenhagen2006} recently interpreting their measurements as
a B0Ve star. 
The optical spectra from ~\citet{levenhagen2004} do not show any
evidence for binarity.

The unequivocal identification of the X-ray source \wga\
\citep{white1994a,white1994b} with \hdstar\ had
previously been thwarted by the lack of sufficient spatial
resolution in the X-ray band to eliminate source confusion in the
center of NGC~5281 as well as
the proximity on the sky of a young supernova remnant (SNR) \gton\ for
which \wga\ could potentially have been a pulsar-wind-nebula (PWN).  
Herein we present new \chandra X-ray and \magell H$\alpha$
observations with the spatial resolution to unambiguously establish
the correspondence between \wga\ and \hdstar\ and discuss the deep 
X-ray spectrum of \hdstar\ taken with \xmm that even at CCD spectral 
resolution reveals emission lines from highly ionized Fe indicative of
a hot thermal plasma. 

\section{Observations}

To obtain an accurate and more finely constrained X-ray source
position we used an archival \chandra observation of the field.  
On 2004 December 26, NGC 5281 was observed for 15~ks by the backside
illuminated chip (S3) of the Advanced CCD Imaging Spectrometer of
\chandra (OBSID:4554). 
The data were reduced and preliminarily analyzed using the
pipeline developed for the \chandra star formation archive
ANCHORS 
\citep{spitzbart2005}. With 1496 net photons from \wga, the 
\chandra spectrum suffers mildly from pile-up ($\sim$12\%).  
In addition to the known bright X-ray source, 
45 other point sources (19 with more than 20 net counts each)
were detected, a number of which are coincident with other members of
the open cluster. This field also contains a large portion of the supernova
remnant (SNR) \gton, believed to lie a few parsecs behind NGC 5281. 
The radio source ATCA J134649-625235, identified by \citet{gaensler1998}
as a candidate pulsar-wind-nebula (PWN) associated with 
this SNR is detected as an X-ray source, with 23
counts 3\arcmin\ away from \hdstar.

For the purposes of accurate astrometry and to reduce source 
confusion a 5~s narrow-band H$\alpha$ image of the central
2.3\arcmin\ of NGC 5821 was taken
with the MagIC CCD camera on the 6.5~m Clay Telescope of the \magell
Observatory at Las Campanas, Chile on 2004 February 17~UT. 
The excellent $<$0.45\arcsec\ seeing and
0.069\arcsec\ 
pixel scale allowed us to register the field relative to the Guide Star
Catalogue-II to an accuracy of  $<$0.75\arcsec\ with 30 matched stars using
WCSTools \citep{mink2002}.
The resulting coordinates for \hdstar\ were 13$^{\mathrm
  h}$46$^{\mathrm m}$32\fs56,-62\degr55\arcmin24\farcs06 (J2000).
We can also conservatively rule out any bright optical companion
further than 0.75\arcsec\ away simply from the lack of any
non-instrumental distortions in the profile of \hdstar\ ($\sim$0.006~pc
at the distance to NGC 5281).  

\begin{figure}
\includegraphics[width=3.3in]{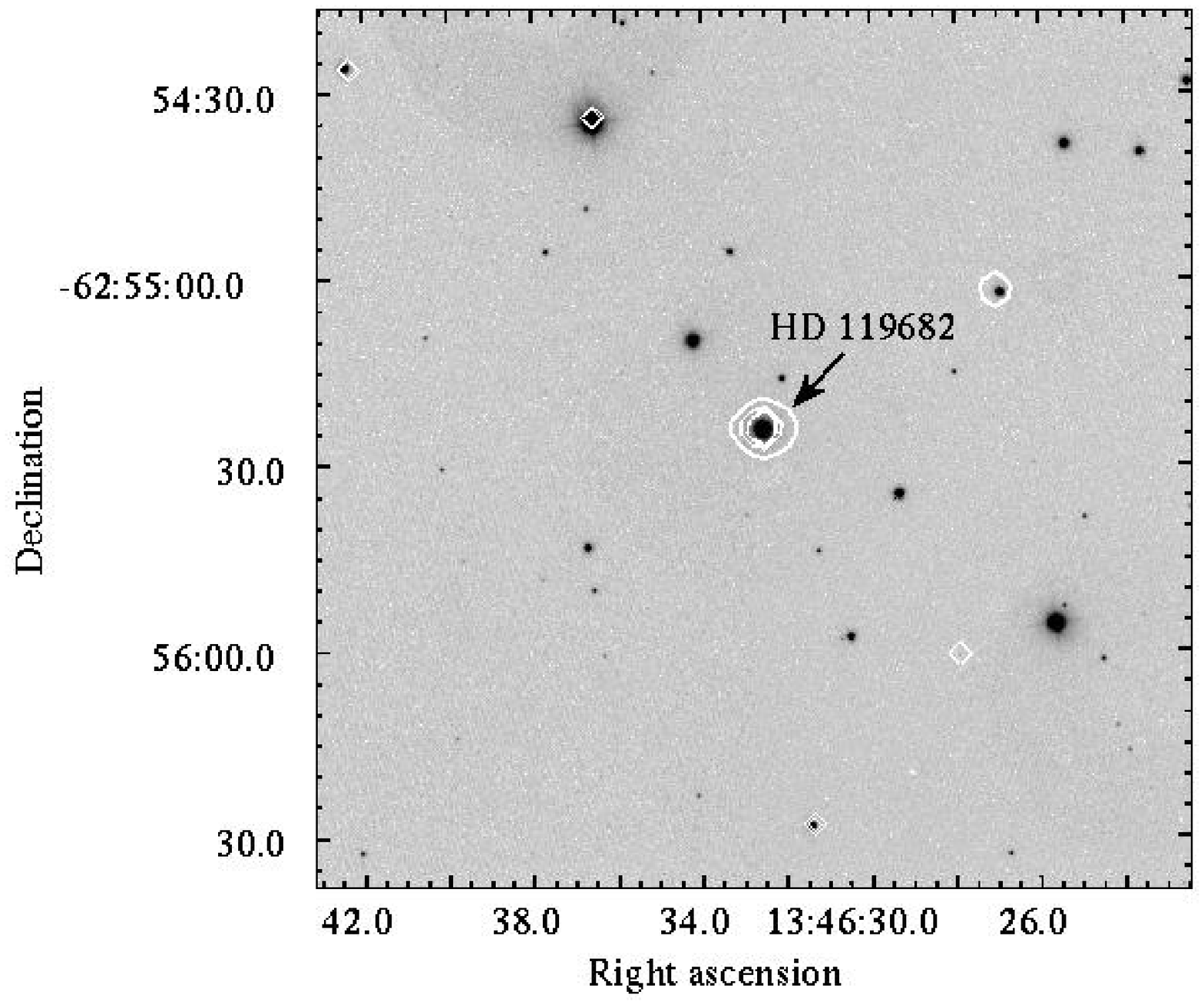}
\caption{
  H$\alpha$ image of open cluster NGC~5281 with \chandra
  X-ray contours overlaid. The H$\alpha$ image was taken with the
  MagIC camera at \magell to better determine the positions of the
  bright stars. The contours shown are 1, 5, and 50 counts
  after adaptive smoothing. \wga\ is clearly identified as the O9
  emission line star \hdstar . (The radio source ATCA J134649-625235
  and its \chandra X-ray counterpart lie 3\arcmin\ to the northwest of
  this image.)  
\label{Halpha}}
\end{figure}

A 39.9~ks \xmm\ observation of SNR \gton\ was taken on 2001 August 28,
OBSID:0087940201 for which NGC~5281 was in the field, 6.5\arcmin\
off-axis.
Three significant 
background flares occurred during the observation affecting EPIC-PN
slightly more than the MOS instruments, resulting in filtered
exposure times of 21.9 ks and 27.0 ks for PN and MOS respectively.
The events were filtered using the standard criteria and
XMM-SAS tools. For spectra from the MOS instruments, patterns between
0 and 12 were kept. For the PN two spectra were extracted, one with
only the single pixel events (pattern=0) and one for the multiple
pixel events (patterns 1 to 4). 
The single pixel response is slightly better known and we can compare
the single and multiple event spectra for signs of pile-up.
Although extremely bright, we find that \wga\ is not piled-up in any
of the \xmm instruments. 
A 30\arcsec\ radius region centered at the \chandra source position
was used to extract the events from \wga .
With 4068, 2677, 2696, and 2880 counts in EPIC PN single, EPIC PN
multiple, MOS1, and MOS2, respectively, background subtraction 
is essentially negligible for \wga , 
reducing the count-rates by less than 5\% .

\section{ X-ray Properties of \hdstar } 

\begin{figure}
\includegraphics[width=3.3in]{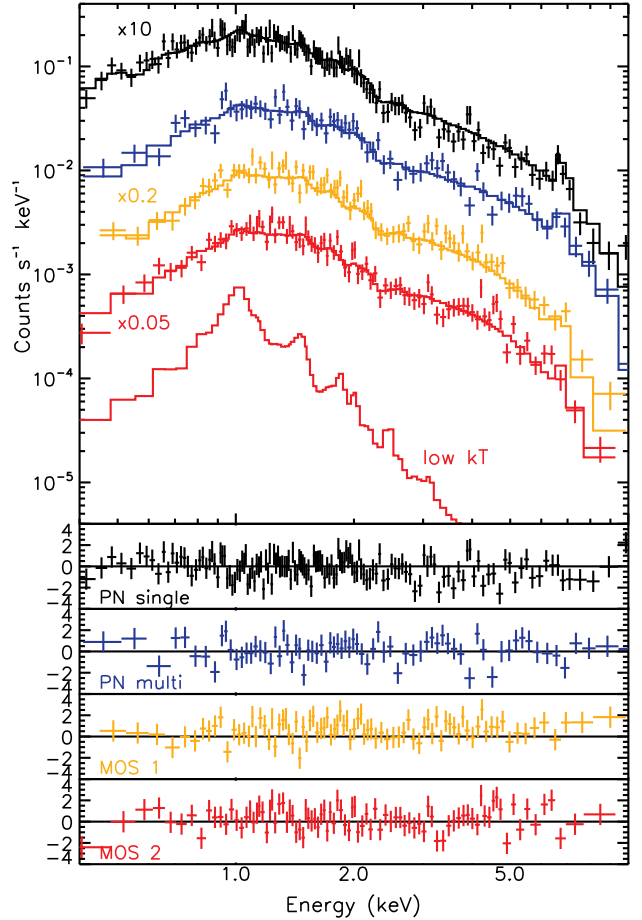}
\caption{\xmm spectra of \wga , from top to bottom, the single event
  PN spectrum ($\times 10$), the multiple event PN spectrum, MOS
  1($\times 0.2$) and MOS2($\times 0.05$). The model shown includes
  two temperature components. The contribution
  of the second cooler component is illustrated for the MOS2 spectrum
  (marked ``low kT''). Fit residuals ($\delta \chi ^{2}$) for each
  instrument are shown below the spectra.
\label{xmm_spec}}
\end{figure}

\asca observations showed that \wga\ had a low
column density more consistent with the distance to the open cluster
NGC~5281 than SNR \gton\ \citep{rakowski2001}.  The \chandra ACIS-S
observations show \wga\ to be an unresolved source 
coincident with \hdstar\ (Figure \ref{Halpha}). 
The flux has remained relatively constant (within 30\%) across 
the \asca, \xmm, and \chandra 
observations in 1999 April, 2001 August, and 2004 December. 
No periodicities were found from the power spectra
of any of the light-curves of \wga . However the light-curves do show
evidence of moderate variability at the 3$\sigma$ level.
The longer \xmm observation and superior high-energy response of
EPIC-PN reveal strong emission lines of Fe~XXV and Fe~XXVI just below 7~keV.
Weaker thermal emission lines
are apparent between 1 and 3 keV (Figure \ref{xmm_spec}).

\begin{deluxetable}{lcc}
\tabletypesize{\small}
\tablecaption{Thermal (Raymond-Smith) Model Parameters \label{xmmtable}}
\tablehead{
\colhead{Parameter} & \multicolumn{2}{c}{Value} \\
\colhead{}   & \colhead{1 $kT$} & \colhead{2 $kT$} }
\startdata
$N_{\mathrm H}$ ($10^{21}$ atoms cm$^{-2}$)
&  1.8$\pm$0.1         &  1.9$\pm$0.1           \\
abundance ($\times$ solar)
&0.39$\pm$0.13      &  0.40$\pm$0.17          \\
$kT$(soft)(keV)
& \nodata            &  1.07$\pm$0.16          \\
flux($10^{-12}$ ergs cm$^{-2}$ s$^{-1}$)\tablenotemark{a}
&  \nodata            &  0.08$\pm 0.04$   \\  
$kT$(hot) (keV)
& 8.1$^{+0.9}_{-0.8}$  &  10.4$^{+2.3}_{-1.6}$   \\
flux($10^{-12}$ ergs cm$^{-2}$ s$^{-1}$)\tablenotemark{a}
& 1.81$\pm$0.01   &  $1.74\pm0.03$  \\ 
$\chi^{2}$(d.o.f.)
&  450.56(390)        & 422.16(388)             
\enddata
\tablenotetext{a}{X-ray flux from 0.5--8.0 keV}
\end{deluxetable}

The \xmm spectrum was modeled by a Raymond-Smith thermal plasma 
model \citep{raymondsmith} with one or two temperature ($kT$)
components absorbed by the intervening column density $N_{\mathrm
  H}$. The metal abundance of the plasma was allowed to vary freely. 
The best-fit parameters, fluxes, 90\% confidence intervals,
and $\chi ^{2}$ values are given in Table \ref{xmmtable}. 
We confirm the low value of $N_{\mathrm H}$ and find a $kT$ of 
$8.1^{+0.9}_{-0.8}$~keV given a single component.
A second cooler component, $kT\sim1$~keV, significantly improves the
fit ($F$-test null-hypothesis probability $< 3.3 \times 10^{-6}$). 
However the flux in this component is $<5$\% of the flux in the hot component.
Based on these spectra we conclude that the emission is dominated by a
extremely hot thermal plasma with $kT=10.4^{+2.3}_{-1.6}$~keV.   
At a distance to NGC 5281 of 1580 pc \citep{paunzen2001} the
measured $N_{\mathrm H}$ and fluxes imply
a total X-ray luminosity of $6.2\pm0.1 \times 10^{32}$ \ergsec. 
While a power-law model would of course fit the electron
bremsstrahlung dominated spectrum, the feature just below 7~keV is
well described by \ion{Fe}{25} and \ion{Fe}{26} line emission in a
thermal plasma at the temperature determined by the continuum. 

\section{Discussion}

Using the excellent spatial resolution of \chandra and optical images
from \magell we were able to identify the X-ray source \wga\, thought
to be a possible PWN at the center of SNR 309.2-0.6, as the O9.7e
massive star \hdstar.
The region itself is rather complex. 
The line of sight hosts an extended radio
structure~\citep{gaensler1998} at a distance of about 5 kpc which
appears positionally coincident with a second possibly extended 
X-ray source from the \ros All Sky Survey (RASS)~\citep{voges1999} 1RXS
J134651.1-624843.
Follow-up \asca observations 
detected strong emission-lines from the extended structure confirming
its identification as a young ejecta-dominated SNR~\citep{rakowski2001}.  
 As for the bright point source
\wga , \ros contours already suggested the possibility of an association 
with the foreground cluster NGC 5281 and its third brightest star
\hdstar\ \citep{gaensler1998}. However, its high X-ray brightness
as well as a measured spectral index from the \asca observation 
were equally consistent with a pulsar or synchrotron nebula
interpretation \citep{rakowski2001}. The low absorbing column density
favored an origin in NGC 5281.

The better than 1\arcsec\ agreement between the \chandra source
position and the optical position of \hdstar\ dispels any doubts about
the identification of \wga\ with \hdstar. 
Both the optical and X-ray properties of \hdstar\ are
reminiscent of $\gamma$ Cas where a recent X-ray gratings study concluded that
the Be-star itself is the likely source of X-ray emission \citep{smith2004}. 
First, \hdstar\ has been classified as a very
late O or very early B emission line star, similar in type to
$\gamma$ Cas (B0.5e).  
Second, the likeness between the X-ray properties of \hdstar\ and $\gamma$
Cas is striking. Our spectral analysis produced a two component fit
that features a 1 keV soft component and a dominant hard
component of about 10 keV.  
Significant \ion{Fe}{25} and \ion{Fe}{26} line emission highlights the
hot thermal  
origin of the spectrum in \hdstar. Likewise $\gamma$ Cas exhibits a very
hot emission component up to $kT=12$~keV along with some levels of
soft emission. 
In both stars the hard emission dominates, with the soft
component contributing only 5\% of the flux in \hdstar\ and 14\% in
$\gamma$~Cas~\citep{smith2004}, which
to our knowledge is unique in stellar systems.  
Furthermore, the measured luminosities are quite similar,
$L_{X}=6.2\times10^{32}$ \ergsec, $log(L_{X}/L_{bol})=-5.4$ for
\hdstar\ compared to  $L_{X}=5.1\times10^{32}$ \ergsec \citep{kubo1998},
$log(L_{X}/L_{bol})=-5.4$ for $\gamma$ Cas using $L_{bol}$ from
\citet{hamann1992}.
That the emission from \hdstar\ is analogous
to $\gamma$~Cas, seems by far the most likely possibility.

By contrast, the colliding wind scenario would have to involve another
similarly massive companion with extremely strong winds, 
no evidence for which is seen
optically in imaging or spectroscopy. Furthermore the spectral type of
\hdstar\ is O9.7e or later. 
For such late type stars the energetics for colliding winds as the
origin of the hard X-ray emission are even more difficult to justify. 
A recent study by~\citet{schulz2006} involving the massive triple
$\theta^2$ Ori A, which contains an O-star of about similar type,
showed that even for the 
existence a close massive companion, the wind densities as well as
the involved kinetics do not favor a colliding wind scenario.
\hdstar\ should be even less favorable as optical spectra do not
indicate a massive companion.

The possibility that the origin of the high-energy emission 
from \hdstar\ is a magnetically confined wind seems a viable alternative. 
However, conditions that lead to a confinement scenario, as modeled for the 
case of $\theta^1$ Ori C~\citep{gagne2005}, seem a bit unrealistic here
given the extremely high plasma temperatures of \hdstar.
Furthermore most magnetic activity of such kind has
been observed in much younger OB stars~\citep{schulz2003}. Quite
recently it was concluded that the field in young $\tau$ Sco, another
young early type star that shows hard X-ray emission \citep{cohen2003},
is likely
of fossil origin rather than some intrinsic dynamo
activity~\citep{donati2006}. The answer thus has to be more complex.  

The remaining possibility points to the quite enigmatic case of
$\gamma$ Cas. 
For the moment, though, with only a CCD-resolution spectrum we cannot go much
beyond a discussion about the similarities between $\gamma$ Cas and \hdstar.
However, the case may bear implications with respect to the nature
of the X-rays in both sources. \hdstar\ is quite well identified
and neither optical spectra~\citep{levenhagen2004} nor our high
spatial resolution optical imaging yet show any evidence for a companion. 
Even though the high resolution
spectra of $\gamma$ Cas provided additional evidence that the observed
X-ray emission is likely not due to a possible white dwarf companion,
that hypothesis could not be entirely ruled out because of
some resemblance between the X-ray emission from some cataclysmic
variables and the HETG spectra presented by~\citet{smith2004} 
More recently~\citet{smith2006} reported on another Be-star (B1e), HD
110432 (BZ Crucis), which shares similarly peculiar X-ray and optical
characteristics with $\gamma$ Cas. The case of \hdstar\
thus seems to be another candidate of ``$\gamma$ Cas analogs'' with the
strong notion that the hard X-ray emission emanates from a single OB
emission line star. 

\acknowledgements
C.E.R. was supported during this work by NASA Grant NAG5-9281.
S.J.W. acknowledges support from CXC guest investigator grant
GO5-6002A (ANCHORS) and NASA contract  NAS8 39073 (CXC).

{\it Facilities:} \facility{Magellan:Clay (MagIC)}, \facility{CXO
  (ACIS)}, \facility{XMM (EPIC)}

\end{document}